# SOCIAL MEDIA
# AS POLITICAL PARTY CAMPAIGN IN INDONESIA

**Leon Andretti Abdillah**
**Bina Darma University**
Jln. AhmadYani No. 12, Plaju, Palembang
E-mail : **leon.abdillah@yahoo.com**

*Abstract:* *Social media as a trend in the Internet is now used as a medium for political campaigns. Author explores the advantages and social media implementation of any political party in Indonesia legislative elections 2014. Author visited and analyzed social media used by the contestants, such as: Facebook, and Twitter. Author collected data from social media until the end of April 2014. This article discusses the use of social media by political parties and their features. The results of this study indicate that social media are: 1) effective tool for current and future political campaigns, 2) reach the voters and supporters instantly, 3) used by Political parties to show their logo/icon, and 4) last but not least quick count results also show that political parties which using social media as part of their campaigns won the legislative elections.*

*Abstrak:* *Media sosial sebagai tren di internet saat ini digunakan sebagai media kampanye politik. Penulis mengeksplor keuntungan dan implementasi media sosial dari partai politik di pemilu legislatif Indonesia 2014. Penulis mengunjungi dan menganalisis media sosial yang digunakan para kontestan, seperti: Facebook dan Twitter. Penulis mengumpulkan data dari media sosial sampai dengan akhir April 2014. Artikel ini mendiskusikan sosial media yang digunakan oleh partai politik beserta fitur-fiturnya. Hasil penelitian ini menunjukkan bahwa media sosial adalah: 1) alat yang efektif untuk kampanye politik saat ini dan masa depan, 2) menggapai pemilih dan pendukung langsung, 3) yang digunakan oleh partai-partai politik untuk menunjukkan logo/icon mereka, dan 4) hasil hitung cepat juga menunjukkan bahwa partai-partai politik yang menggunakan media sosial sebagai bagian dari kampanye mereka memenangkan pemilu legislatif.*

*Kata kunci: Dampak media sosial, pemilu legislatif Indonesia, presentasi partai politik.*

## 1. INTRODUCTION

One of the most popular internet application nowadays are social media sites. These social media applications grow significantly and attract many concerns from online users. At the moment, social media have been used for personal communication, education (Abdillah, 2013), promotion (Rahadi & Abdillah, 2013), and knowledge and information sharing (Abdillah, 2014). The rapid development of online social networks has tremendously changed the way of people to communicate with each other (Bi, Qin, & Huang, 2008). This article will cover the topic of social media as political party campaign on involving citizens in the democratic activity like general legislative elections or presidential campaigns.

Every democratic country would have done a good election to select members of council or parliament, as well as to the president. Normally, in the campaign periods every political party will promote their candidate using various of media.

Usually before the elections, the candidate will conduct a promotion or campaign that contains a call to pick him. Various media have been used to support the purposes of the campaign. Along with the development of information technology is increasingly widespread and rapid, the candidates also campaigned increasingly familiar with using information technology.

The most current information technology application to promote political campaigns is social media. This happends because information networks not easily controlled by the state and coordination tools that are already embedded in trusted networks of family and friends (Howard & Hussain, 2011). Organizations such as political parties are trying to keep up with this changing environment (Effing, van Hillegersberg, & Huibers, 2011). Another reason is based on one common characteristic among social media sites is that they tend to be free and are therefore widely accessible across socioeconomic classes (Joseph, 2012). Last but not least, adding new media to old electoral politics will entice new and younger voters to greater participation (Xenos & Foot, 2008), because there are relationships between Facebook use and students' life satisfaction, social trust, civic engagement, and political participation (Valenzuela, Park, & Kee, 2009).

Author itself has been used social media for supporting student learning environment (Abdillah, 2013), promotion media (Rahadi & Abdillah, 2013), and knowledge sharing (Abdillah, 2014) in higher education institution. Other article report how social media is emerging as an important technology for disaster response (Yates & Paquette, 2011).

And now social media is affecting political campaigns (Smith, 2011), including young adults engagements (Baumgartner & Morris, 2010). Educated and well inform people less trust to billboards or banners, but they have more confidence or rather believe in the words of friends or colleagues in social media (Sugiarto, 2014).

One of the most phenomenon is Barack Obama's campaign in 2008. The successful use of social media in the US presidential campaign of Barack Obama (Tumasjan, Sprenger, Sandner, & Welpe, 2010) has established Twitter, Facebook, MySpace, and other social media as integral parts of the political campaign toolbox and how they have affected users' political attitudes and behaviors (Zhang, Johnson, Seltzer, & Bichard, 2010). Another success story is from Indonesia, Jokowi & Ahok, new Governor and deputy as winners of Jakarta Governor Election in the 2012s suggest political marketing strategy is an effective key to success (Ediraras, Rahayu, Natalina, & Widya, 2013). On of their political branding in governor election campaign is twitter social media (Wulan, Suryadi, & Dwi Prasetyo, 2014). Political communication alse uses blog hyperlinks that including political party, activist groups, and individuals (Rosen, Barnett, & Kim, 2010).

In this article, author would like to discuss about successful story in social media based political campaign in Indonesia.

Per March 2014, Indonesia was the fourth larger facebook users (SocialBakers, 2014) after USA, India, and Brazil. Facebook users of Indonesia is dominanted by young adults (19-24 years) people followed by Adults (25-34 years), see figure 1.

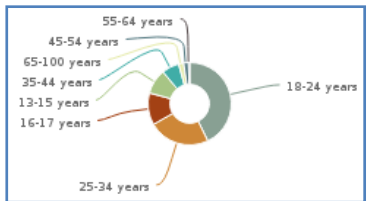

**Figure 1. Facebook users based on age**

If we check the users of facebook based on gender, male are dominant users for Indonesian facebook users (SocialBakers, 2014), see figure 2.

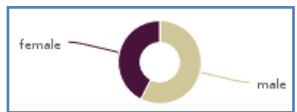

**Figure 2. Facebook users based on gender**

The rest of this paper will cover research methods, results and discussions, and conclusions.

## 2. RESEARCH METHODS

This article will observe the online features of political parties' social media, such as: 1) Facebook, and 2) Twitter. The author explored the political parties' social media to check their activities. Author also gathered some valuable information from popular new websites such as: 1) sindo.com, 2) kompas.com, and 3) liputan6.com.

### 2.1 Political Parties

In the five-yearly Indonesia general election in 2014 will be followed by twelve national political parties (Table 1) and three local political parties in Aceh.

In this article, author will not discuss three others local political parties in Aceh. Focus analysis and discussion only for twelve national political parties.

**Table 1. Polical Parties in Indonesia General Elections 2014**

| No | Political Party | Head, General Secretary, Treasury, & website | Logo |
|---|---|---|---|
| 1 | Partai Nasional Demokrat (Nasdem) | Surya Paloh, Patrice Rio Capella, Frankie Turtan http://www.partainasdem.org/ | |
| 2 | Partai Kebangkitan Bangsa (PKB) | A.Muhaimin Iskandar, Imam Nahrawi, Bachrudin Nasori http://www.dpp.pkb.or.id/ | |
| 3 | Partai Keadilan Sejahtera (PKS) | Muhammad Anis Matta, Muhamad Taufik Ridlo, Mahfudz Abdurrahman http://www.pks.or.id/ | |
| 4 | Partai Demokrasi Indonesia Perjuangan (PDIP) | Megawati Soekarno Putri, Tjahjo Kumolo, Olly Dondokambey http://www.pdiperjuangan.or.id/ | |
| 5 | Partai Golongan Karya (Golkar) | Aburizal Bakrie, Idrus Marham, Setya Novanto http://partaigolkar.or.id/ | |
| 6 | Partai Indonesia Raya (Gerindra) | Suhardi, Ahmad Muzani, Thomas A. Muliatna Djiwandono http://partaigerindra.or.id/ | |
| 7 | Partai Demokrat (PD) | Susilo Bambang Yudhoyono, Edhie Baskoro Yudhoyono, Handoyo Mulyadi http://www.demokrat.or.id/ | |
| 8 | Partai Amanat Nasional (PAN) | M. Hatta Rajasa, Taufik Kurniawan, Jon Erizal http://pan.or.id/ | |
| 9 | Partai Persatuan Pembangunan (PPP) | Suryadharma Ali, Romahurmuziy, Mahmud Yunus http://ppp.or.id/index.html | |
| 10 | Partai Hati Nurani Rakyat (Hanura) | Wiranto, Dossy Iskandar Prasetyo, Bambang Sudjagad http://hanura.com/10/ | |
| 11 | Partai Bulan Bintang (PBB) | MS. Kaban, B.M. Wibowo, Sarinandhe Djibran http://bulan-bintang.org/ | |
| 12 | Partai Keadilan dan Persatuan Indonesia (PKPI) | Sutiyoso, Lukman F. Mokoginta, Linda Setiawati http://pkpi.or.id/ | |

Source: Komisi Pemilihan Umum (KPU, 2014)

Information about every national poltical party's website, facebook, twitter, etc will be discussed in next section.

## 2.2 Political Parties' Social Media

Until April 2014, eleven of twelve political parties use Facebook, and all political parties uses Twitter. Table 2 shows social media account of political parties in Indonesia.

**Table 2. Political Parties Social Media**

| No | Political Party | Facebook Page | Twitter |
|---|---|---|---|
| 1 | P.Nasdem | - | @NasDem |
| 2 | PKB | pkb2pkb | @PKB_News_Online |
| 3 | PKS | HumasPartaiKeadilanSejahtera | @PKSejahtera |
| 4 | PDIP | DPP.PDI.Perjuangan | @PDI_Perjuangan |
| 5 | P.Golkar | DPPPGolkar | @Golkar2014 |
| 6 | P.Gerindra | gerindra | @Gerindra |
| 7 | P.Demokrat | pdemokrat | @PDemokrat |
| 8 | PAN | amanatnasional | @official_PAN |
| 9 | PPP | pppdpp | @DPP_PPP |
| 10 | P.Hanura | hanura.official | @hanura_official |
| 11 | PBB | DPP-Partai-Bulan-Bintang-wwwbulan-bintangorg/114716555303039 | @DPPBulanBintang |
| 12 | PKPI | PKPI.MediaCenter | @sobatbangyos |

## 3. RESULTS AND DISCUSSIONS

In this section, author would like to show the results from quick counts institutions, then followed by information about political party's social media.

## 3.1 Quick Counts Results

Quick count have been used for several general elections in Indonesia. Table 3 shows the quick count results from three survey institutions: 1) CSIS-Cyrus (Gunawan, 2014), 2) Litbang Kompas (LITBANG KOMPAS, 2014), 3) IRC-Sindo (Prawira, 2014), and 4) Lingkar Survei Indonesia-LSI (Rastika, 2014).

Based on quick counts result, we have the temporer winner of the general election in Indonesia: 1) PDIP, 2) PGolkar, 3) PGerindra, 4) PDemokrat, 5) PKB, 6) PAN, 7) PNasdem, 8) PPP, 9) PKS, 10) PHanura, 11) PBB, 12) PKPI.

**Table 3. Quick Count Results in Indonesia General Election 2014**

| No | Political Party | CSIS-Cyrus Network | Litbang Kompas | IRC-Sindo | Lingkar Survei Indonesia (LSI) |
|---|---|---|---|---|---|
| 1 | P.Nasdem | 6.9 % | 6.71 % | 6.41 % | 6.35 % |
| 2 | PKB | 9.2 % | 9.12 % | 9.51 % | 9.30 % |
| 3 | PKS | 6.9 % | 6.99 % | 7.11 % | 6.46 % |
| 4 | PDIP | 18.9 % | 19.24 % | 18.96 % | 19.65 % |
| 5 | P.Golkar | 14.3 % | 15.01 % | 14.90 % | 14.95 % |
| 6 | P.Gerindra | 11.8 % | 11.77 % | 11.90 % | 11.79 % |
| 7 | P.Demokrat | 9.7 % | 9.43 % | 9.20 % | 9.68 % |
| 8 | PAN | 7.5 % | 7.51 % | 7.06 % | 7.52 % |
| 9 | PPP | 6.7 % | 6.68 % | 6.81 % | 6.95 % |
| 10 | P.Hanura | 5.4 % | 5.10 % | 5.34 % | 5.21 % |
| 11 | PBB | 1.6 % | 1.50 % | 1.61 % | 1.33 % |
| 12 | PKPI | 1.1 % | 0.95% | 1.18 % | 0.97 % |

April 2014

Unfortunately there is no political party get 20% or more to support their own president candidate. Author classifies three big political parties (PDIP, PGolkar, PGerindra), three medium political parties (PDemokrat, PKB, PAN), four small political parties (PNasdem, PKS, PPP, PHanura), and two political parties which threatened not to go to parliament.

## 3.2 National Political Parties' Facebook Page

The Facebook connectivity help the group to build a political party to further back up their main figure in the forecast Presidential candidacy (Murti, 2013). Table 4 shows the popularity of every political partiy in Facebook social media per April 2014.

**Table 4. Political Parties Facebook's Like**

| No | Political Party | Facebook Page | Like |
|----|-----------------|---------------|------|
| 1  | P.Nasdem  | - | - |
| 2  | PKB       | pkb2pkb | 6.164 K |
| 3  | PKS       | HumasPartaiKeadilanSejahtera | 40.073 K |
| 4  | PDIP      | DPP.PDI.Perjuangan | 319.000 K |
| 5  | P.Golkar  | DPPPGolkar | 4.355 K |
| 6  | P.Gerindra| gerindra | 2.500 M |
| 7  | P.Demokrat| pdemokrat | 25.075 K |
| 8  | PAN       | amanatnasional | 38.228 K |
| 9  | PPP       | pppdpp | 3.388 K |
| 10 | P.Hanura  | hanura.official | 562.000 K |
| 11 | PBB       | DPP-Partai-Bulan-Bintang-wwwbulan-bintangorg/114716555303039 | 2.198 K |
| 12 | PKPI      | PKPI.MediaCenter | 4.410 K |

April 2014

Based on Political Parties' Facebook Page Likes, there are three the most popular political parties in facebook: 1) PGerindra, 2) PHanura, and 3) PDIP. Among those three political parties, PGerindra and PDIP are the two top political parties at the moment.

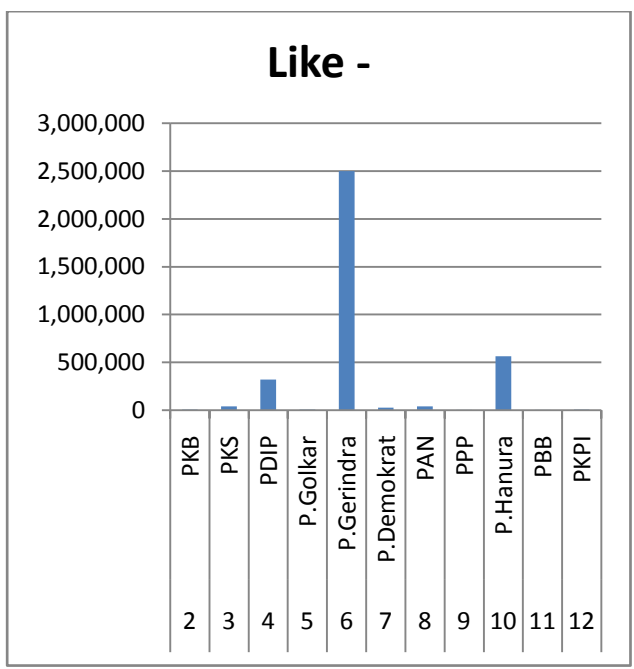

**Figure 3. Political Parties' Facebook Page Like's Statistics**

### 3.3 National Political Parties' Twitter

All of national political parties have official website (Table 1) and Twitter account (Table 2). Almost twitter accounts from all political parties included the abreviation of their political party's name except PKPI which use @sobatbangyos. Table 4 shows political partiy's Facebook's like.

Table 5 shows the numbers of tweets, following, and followers from every political parties (in Killo). All of political parties have twitter account.

**Table 5. Political Parties Twitter's Tweets, Following, and Followers**

| No | Political Party | Tweets | Following | Followers |
|----|-----------------|--------|-----------|-----------|
| 1  | P.Nasdem    | 17.600 K | 0.669 K | 20.900 K |
| 2  | PKB         | 2.898 K  | 1.705 K | 4.050 K  |
| 3  | PKS         | 18.900 K | 0.275 K | 105.000 K |
| 4  | PDIP        | 21.300 K | 0.658 K | 58.400 K |
| 5  | P.Golkar    | 9.680 K  | 0.493 K | 2.329 K  |
| 6  | P.Gerindra  | 47.600 K | 2.160 K | 143.000 K |
| 7  | P.Demokrat  | 4.008 K  | 0.870 K | 18.200 K |
| 8  | PAN         | 6.022 K  | 0.605 K | 4.745 K  |
| 9  | PPP         | 4.424 K  | 0.076 K | 2.953 K  |
| 10 | P.Hanura    | 1.256 K  | 0.030 K | 1.866 K  |
| 11 | PBB         | 0.172 K  | 0.014 K | 0.774 K  |
| 12 | PKPI        | 2.301 K  | 1.376 K | 1.364 K  |

April 2014

In this article, author will analyze twitter features of every political party based on 1) tweets, 2) following, and 3) followers.

Figure 4.a shows the numbers of tweets of every political party. P.Gerindra was the most informative political party with 47,6 K tweets followed by P.DIP and PKS.

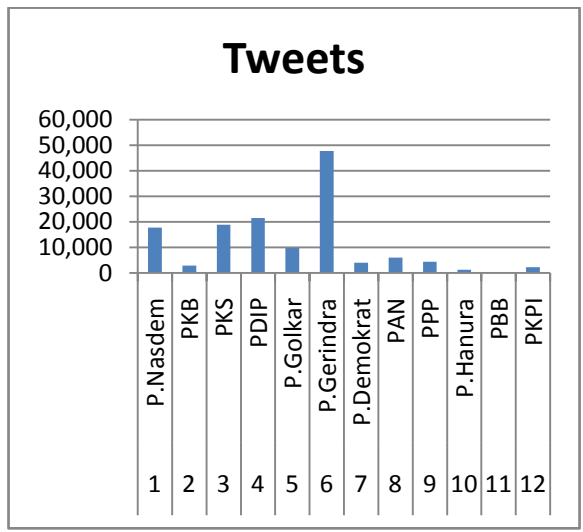

**Figure 4.a. Political Parties' Twitters Tweets**

Figure 4.b shows the numbers of following of every political party. PGerindra was the most kindest political party that has been following 2,16K others Twitter accounts followed by PKB and PKPI.

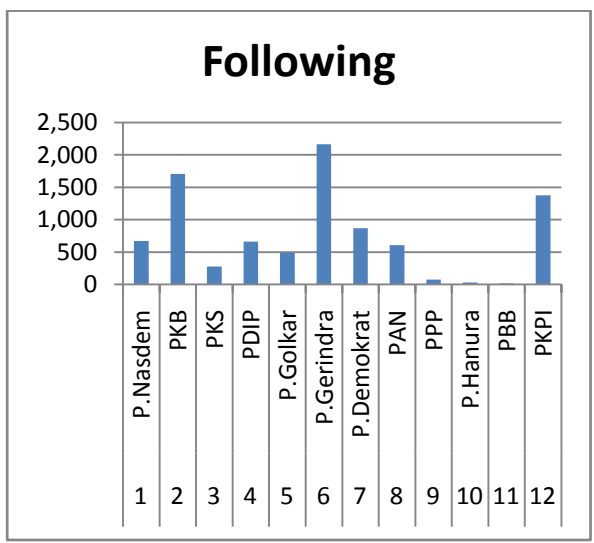

**Figure 4.b. Political Parties' Twitters Following**

Table 4.c shows the numbers of political parties' twitters followers. Based on that statistic, P.Gerindra has the largest number of followers followed by PKS and PDIP.

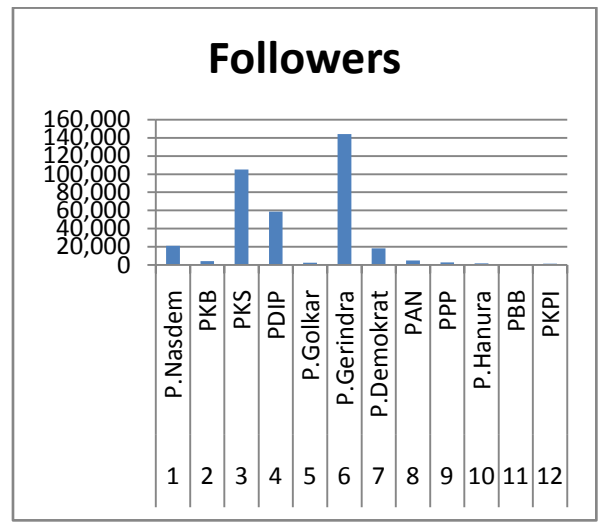

**Figure 4.c. Political Parties' Twitters Followers**

Three political parties that get the most twitter followers are: 1) PGerindra, 2) PKS, and 3) PNasdem.

### 3.4 The Comparison Between Social Media Data and Political Party Rank based on Quick Count Results

Based on the result discussed above, social media shows data into political party to get the trust from their voters, participants, and volunteers. Table 6 summarize social media used by political parties in Indonesian legislative general elections 2014.

Based on the comparison between quick count results versus social media data, two political parties of PGerindra and PDIP succeed to collect trust from voters. These two political parties are the big three winner for 2014 legislative general elections in 2014. Another winner is PGolkar stable to keep their loyal participants and need less effort to reach new voters via social media. But if PGolkar also increase their concerns in utilizing the power of socal media for their campaigns, their result could be much better.

**Table 6. The Comparison of Quick Counts Results versus Social Media Data**

| No | Quick counts | Facebook | Twitter |
|---|---|---|---|
| 1 | PDIP | PGerindra | PGerindra |
| 2 | PGolkar | PHanura | PKS |
| 3 | PGerindra | PDIP | PDIP |
| 4 | PDemokrat | PKS | PNasDem |
| 5 | PKB | PAN | PDemokrat |
| 6 | PAN | PDemokrat | PAN |
| 7 | PNasDem | PKB | PKB |
| 8 | PPP | PKPI | PPP |
| 9 | PKS | PGolkar | PGolkar |
| 10 | PHanura | PPP | PHanura |
| 11 | PBB | PBB | PKPI |
| 12 | PKPI | PBB | PBB |

Another condition is faced by PKS. This political party has been used all popular social media, unfortunately PKS is not succeed to get better voices. It is due to other causes that are not related to the use of social media, and author does not discuss it in this article.

## 3.5 Social Media as Political Party Brand

Social Media has ability to promote a political party's image over the world easily. Every political party provides social media for their loyal audiences, and to get more online active persons.

The result from SocialBakers show that political parties or political figures pick the high interest from social media community.

Based on figure 5, Prabowo Subianto the chairman of PGerindra and his party (PGerindra) successfully perched on top position and number three in social media society. These achievement also followed by the quick count results, as PGerindra occupy the third posisition.

## 3.6 Features Presentation in Political Party's Social Media

Social Media has ability to promote a political party image over the world easily. Every political party provide social media for their loyal audiences, and to get more online active persons. Social media platforms give politicians access to millions of users and offer the capacity to build a sense of camaraderie and connection with a wide constituency (Crawford, 2009).

**Table 7. Political Parties' Facebook Page Views**

| No | PP | Logo | No Urut | Head | Quote, Slogan, TagLine | Others |
|---|---|---|---|---|---|---|
| 1 | P.Nasdem | - | - | | | |
| 2 | PKB | ✓ | ✓ | - | - | |
| 3 | PKS | ✓ | - | ✓ | ✓ | Simpatisan |
| 4 | PDIP | ✓ | - | ✓ | - | Soekarno |
| 5 | P.Golkar | ✓ | - | - | ✓ | - |
| 6 | P.Gerindra | ✓ | - | ✓ | ✓ | - |
| 7 | P.Demokrat | ✓ | ✓ | ✓ | ✓ | - |
| 8 | PAN | ✓ | ✓ | ✓ | ✓ | - |
| 9 | PPP | ✓ | ✓ | - | ✓ | - |
| 10 | P.Hanura | ✓ | ✓ | - | ✓ | KPU |
| 11 | PBB | ✓ | - | ✓ | - | Simpatisan |
| 12 | PKPI | ✓ | ✓ | ✓ | ✓ | Pemilu.com |
| | Persentage | 100% | 50% | 58.3% | 66.67% | 41.67% |

April 2014

Author visited every political parties' Facebook account. Author found that every political party has different way to present their existness via facebook. Table 7 shows the summary of political parties' Facebook pages' features. In 2014 general elections, all of political parties dispaly their logo, 50% display their number, 58% display the head of political party, 66,67% have quote/slogan/tagline, and 41.67% display the third party.

Unfortunately only two political parties that displays its supporters in their social media main pages. The political parties seemed forget that the supporters are one of the most important aspect that they should display in their social media main pages.

## 4. CONCLUSIONS

Based on the facts and discussions above, author would like to summarize the condition of social media related to general legislative elections results in Indonesia as follow:

1) Social media is effective tool for political campaigns. The power of social media has triggers transparancy and support e-democracy around the world. Citizens have ability to choose freely the best legislative candidate to represent them in the parliament.
2) Social media are the current and future media for political campaigns and reach the voters and supporters instantly. Political parties are encourage to provide more professional social media pages over the internet.
3) Social media will create a more successful campaign as well as help create a stronger democracy (Vonderschmitt, 2012). There is no wall for every body to search the best candidate through online social media.
4) Facebook like is the symbols of popularity in Indonesian political athmosphere.
5) Another aspect that makes Facebook the most used social network is being able to create an event (Curran, Morrison, & Mc Cauley, 2012).
6) Political parties' logos is the most common icon found in political parties' social media in Indonesian legislative general elections 2014.
7) Information on Twitter can be aggregated in a meaningful way (Tujasman, Sprenger, Sandner, & Welpe, 2010). Total followers have linear correlation with the voters in real election. In the case of Indonesia, PGerindra and PDIP success to get many voters. Contra conditions are faced by PGolkar and PKS. PGolkar uses less effort in social media. PGolkar still works with traditional media, television, and they success to keep their loyal audiens. Another contra condition is faced by PKS, even they already force all of popular social media for political campaigns, the acquisition constituents that they obtained is lower than it should be.
8) For future reseach, author interested to explore the power of social media in poilitics combined with blogs as digital democracy (Gil de Zúñiga, Veenstra, Vraga, & Shah, 2010). Focus observation involves presidential candidates and communication relationships with prospective voters.

Author also interested to investigate political events not only in Facebook and Twitter but also involve others social media.